\documentclass[twocolumn]{aastex631}

\usepackage{subfigure}

\shorttitle{VLBA Observations of Wandering Black Hole Candidates}
\shortauthors{Sargent et al.}
\graphicspath{{./}{figures/}}

\begin{document}

\title{Wandering Black Hole Candidates in Dwarf Galaxies at VLBI Resolution}

\email{andrew.j.sargent2.civ@us.navy.mil}
\AuthorCollaborationLimit=8

\author[0000-0002-8049-0905]{Andrew J.\ Sargent}
\affiliation{United States Naval Observatory,
3450 Massachusetts Ave., NW,
Washington, DC 20392, USA}
\affiliation{Department of Physics, The George Washington University, 725 21st St. NW, Washington, DC 20052, USA}

\author[0000-0002-4146-1618]{Megan C. Johnson}
\affiliation{United States Naval Observatory,
3450 Massachusetts Ave., NW,
Washington, DC 20392, USA}

\author[0000-0001-7158-614X]{Amy E.\ Reines}
\affiliation{eXtreme Gravity Institute, 
Department of Physics, 
Montana State University, 
Bozeman, MT 59717, USA}

\author[0000-0002-4902-8077]{Nathan J.\ Secrest}
\affiliation{United States Naval Observatory,
3450 Massachusetts Ave., NW,
Washington, DC 20392, USA}

\author[0000-0001-9149-6707]{Alexander J.\ van der Horst}
\affiliation{Department of Physics, The George Washington University, 725 21st St. NW, Washington, DC 20052, USA}
\affiliation{Astronomy, Physics and Statistics Institute of Sciences (APSIS), The George Washington University, Washington, DC 20052, USA
}

\author[0000-0002-8736-2463]{Phil J.\ Cigan}
\affiliation{Department of Physics and Astronomy, George Mason University, MS3F3, 4400 University Drive, Fairfax, VA 22030, USA}
\affiliation{United States Naval Observatory,
3450 Massachusetts Ave., NW,
Washington, DC 20392, USA}

\author[0000-0003-2511-2060]{Jeremy Darling}
\affiliation{Center for Astrophysics and Space Astronomy, Department of Astrophysical and Planetary Sciences, University of Colorado, 389 UCB, Boulder, CO 80309-0389, USA}

\author[0000-0002-5612-3427]{Jenny E.\ Greene}
\affiliation{Department of Astrophysical Sciences, Princeton University, Princeton, NJ 08544, USA}

\begin{abstract}
Thirteen dwarf galaxies have recently been found to host radio-selected accreting massive black hole (MBH) candidates, some of which are ``wandering" in the outskirts of their hosts. We present 9 GHz Very Long Baseline Array (VLBA) observations of these sources at milliarcsecond resolution. Our observations have beam solid angles ${\sim}10^4$ times smaller than the previous Very Large Array (VLA) observations at 9 GHz, with comparable point source sensitivities. We detect milliarcsecond-scale radio sources at the positions of the four VLA sources most distant from the photo-centers of their associated dwarf galaxies. These sources have brightness temperatures of ${>}10^6~\mathrm{K}$, consistent with active galactic nuclei (AGNs), but the significance of their preferential location at large distances ($p$-value~$=0.0014$) favors a background AGN interpretation. The VLBA non-detections toward the other 9 galaxies indicate that the VLA sources are resolved out on scales of tens of milliarcseconds, requiring extended radio emission and lower brightness temperatures consistent with either star formation or radio lobes associated with AGN activity. We explore the star formation explanation by calculating the expected radio emission for these nine VLBA non-detections, finding that about 5 have VLA luminosities that are inconsistent with this scenario. Of the remaining four, two are associated with spectroscopically confirmed AGNs that are consistent with being located at their galaxy photo-centers. There are therefore between 5 and 7 wandering MBH candidates out of the 13 galaxies we observed, although we cannot rule out background AGNs for five of them with the data in hand.
\end{abstract}

\section{Introduction \label{sec:intro}}
In the concordance $\Lambda$CDM cosmology, galaxies like our own are predicted to have formed from the hierarchical coagulation of smaller, gas-rich stellar systems similar to the dwarf irregular and extremely metal-poor dwarf galaxies observed in our local universe \citep[e.g.,][and references therein]{2020NatRP...2...42V}.
The presence of black holes (BHs) with masses larger than $10^{9}$\,$M_\sun$ at high redshifts \citep[the current record holder is a $1.6\times10^9$\,$M_\sun$ BH at $z=7.642$;][]{2021ApJ...907L...1W} is therefore difficult to explain as the result of the growth of normal stellar BH remnants, and various alternative mechanisms have been proposed for the formation of the first massive BH (MBH; $M_\mathrm{BH} > 100$\,$M_\sun$) ``seeds'' \citep[for recent reviews, see][]{2020ARA&A..58..257G,volonteri2021}. Critically, these mechanisms predict differing BH mass distributions and galaxy occupation fractions, something that is, in principle, observable. However, distinguishing between these mechanisms requires searching for MBHs in dwarf galaxies---galaxies typically defined, in this context, as having a stellar mass less than the Large Magellanic Cloud (LMC), ${\sim}10^{9.5}$\,$M_\sun$---because the hierarchical buildup of larger galaxies along with their MBHs erases information about the original MBH seed masses. For a recent review on MBHs in dwarf galaxies, see \citet[][]{reines2022}.

MBHs in dwarf galaxies are smaller than $10^6$\,$M_\sun$, making them difficult to dynamically detect using current instrumentation \citep[e.g.,][]{nguyen2019}. Their small masses mean that when MBHs in dwarf galaxies accrete and become an active galactic nucleus (AGN), their radiative output is generally faint in comparison to star formation in their host galaxies, making them difficult to detect \citep[e.g.,][]{2015ApJ...811...26T}. Consequently, prior to the advent of large, sensitive sky surveys, only two AGNs in dwarf galaxies were known, NGC~4395 and POX~52 \citep{2003ApJ...588L..13F,2004ApJ...607...90B}. The advent of the Sloan Digital Sky Survey (SDSS), which produced ${\sim}10^6$ galaxy spectra, allowed for dedicated searches for low-mass MBHs and dwarf galaxies with MBHs \citep{2007ApJ...670...92G, 2013ApJ...775..116R} that uncovered several hundred, but these objects nonetheless proved to be rare. For example, out of a parent sample of ${\sim}45,000$ dwarf galaxies selected from SDSS data, \citet{2013ApJ...775..116R} find only 136 AGN and AGN-star-forming-composite galaxies selected via emission line ratio diagnostics, a frequency of 0.3\%, which is a factor of ${\sim}60$ less than emission line-selected AGNs selected from SDSS as a whole \citep[e.g.,][]{2003MNRAS.346.1055K}.

A further issue confounding searches for MBHs in dwarf galaxies is that their gravitational potentials are much shallower than in more massive galaxies. This has the effect of allowing MBHs to ``wander'' around dwarf galaxies, with orbital decay to the gravitational center of the galaxy taking timescales comparable to a Hubble time \citep{2019MNRAS.482.2913B, 2021MNRAS.505.5129B}. Consequently, many MBHs will simply not be found at the centers of dwarf galaxies, leading to significant incompleteness in single-aperture spectroscopic surveys such as the SDSS. Confirming the presence of an accreting, wandering MBH is at least as difficult, using current observational capabilities, as confirming those at the centers of their host galaxies, and gravitational wave facilities that will be particularly sensitive to MBHs in the intermediate mass range, such as the Laser Interferometer Space Antenna \citep[LISA;][]{2017arXiv170200786A} are at least a decade away \citep[e.g.,][]{2019arXiv190706482B}. 

Recent work has found candidate wandering MBHs using both high resolution radio observations \citep{2020ApJ...888...36R}, and spatially-resolved spectroscopic observations \citep{2020ApJ...898L..30M}. Both techniques have drawbacks. In the case of spatially-resolved spectroscopy, off-nuclear AGN candidates may in fact be ``light echoes'' from a past period of enhanced AGN activity originating in the center of the galaxy, as is commonly found in larger galaxies. In the case of radio-selected objects, compact, off-nuclear sources may be supernova remnants or other star formation-related phenomena that are not connected to MBH activity, and the probability of a source being a background AGN increases with the square of the distance from the galaxy center. While these possibilities were carefully considered in \citet{2020ApJ...888...36R}, the $0\farcs25$ angular resolution of the Karl G.\ Jansky Very Large Array (VLA) radio continuum observations corresponds to radio sources that may be as large as $\sim100$~pc, given the typical redshifts of the sample, and therefore follow-up radio observations at Very Long Baseline Interferometry (VLBI) resolutions could confirm the presence of compact radio cores powered by MBH accretion.

In this work, we present follow-up Very Long Baseline Array (VLBA) observations of the 13 sources identified in \citet{2020ApJ...888...36R} as radio-selected MBH candidates in dwarf galaxies.  In Section~\ref{sec:sample} we review this sample, and in Section~\ref{sec:observations} we present the VLBA data and its analysis. Our results are given in Section~\ref{sec:results}, with a discussion in Section~\ref{sec:discussion}. Our main conclusions are listed in Section \ref{sec:conclusion}. For consistency with \citet{2020ApJ...888...36R}, we use a flat $\Lambda$CDM cosmology with $\Omega_\mathrm{m}=0.3$ and $H_0 = 73$~km~s$^{-1}$~Mpc$^{-1}$.

\begin{deluxetable*}{clCLLCCCRRC}
\caption{VLBA Targets}
\tablehead{
\colhead{ID} & 
\colhead{Galaxy Name} & 
\colhead{NSA ID} & 
\colhead{R.A.} & 
\colhead{Dec.} & 
\colhead{$z$} & 
\colhead{$\log{M_{\star}}$} & 
\colhead{$M_g$} & 
\colhead{$g-r$} & 
\colhead{$r_{50}$} & 
\colhead{Sérsic $n$} \\  
& & & \colhead{(h m s)} & \colhead{($^\circ~'~''$)} & & \colhead{($M_{\odot}$)} & \colhead{(mag)} & \colhead{(mag)} & \colhead{(kpc)} & }
\startdata
2  & J0019+1507   &  26027 & 00~18~59.99 & +15~07~11.1 & 0.0376 & 8.65 & -18.42 &  0.15 &  1.51 & 0.8\\
6  & J0106+0046   &  23750 & 01~06~07.31 & +00~46~34.3 & 0.0171 & 9.40 & -18.42 &  0.39 &  5.01 & 1.8\\
25 & J0903+4824   &  26634 & 09~03~13.97 & +48~24~13.7 & 0.0272 & 8.83 & -17.46 &  0.36 &  1.64 & 0.7\\
26 & J0906+5610   &  10779 & 09~06~13.77 & +56~10~15.1 & 0.0470 & 9.36 & -18.98 &  0.40 &  1.64 & 4.1\\
28 & J0909+5655   &  12478 & 09~09~08.69 & +56~55~19.7 & 0.0315 & 8.32 & -17.46 &  0.20 &  1.31 & 1.3\\
33 & J0931+5633   &  16467 & 09~31~38.42 & +56~33~19.9 & 0.0494 & 8.34 & -16.72 &  0.31 &  1.15 & 0.6\\
48 & J1027+0112   & 137386 & 10~27~41.38 & +01~12~06.4 & 0.0212 & 7.82 & -15.81 &  0.27 & 10.36 & 0.6\\
64 & J1136+1252   &  66255 & 11~36~48.53 & +12~52~39.9 & 0.0340 & 9.32 & -18.63 &  0.34 &  2.75 & 1.3\\
65 & J1136+2643   & 101782 & 11~36~42.58 & +26~43~35.7 & 0.0331 & 9.24 & -18.27 &  0.38 &  3.40 & 1.3\\
77 & J1200$-$0341 &   3323 & 12~00~58.30 & -03~41~18.5 & 0.0257 & 9.23 & -18.44 &  0.33 &  2.48 & 1.2\\
82 & J1220+3020   & 102751 & 12~20~11.27 & +30~20~08.3 & 0.0269 & 9.37 & -18.21 &  0.35 &  1.09 & 4.2\\
83 & J1226+0815   &  67389 & 12~26~03.64 & +08~25~19.0 & 0.0241 & 9.31 & -17.93 &  0.59 &  1.25 & 4.6\\
92 & J1253$-$0312 &   3602 & 12~53~06.97 & -03~12~58.8 & 0.0221 & 8.60 & -19.99 & -0.53 &  0.51 & 6.0\\
\enddata
\tablecomments{The thirteen targets observed with the VLBA. Global properties are from the NASA-Sloan Atlas (NSA) \texttt{v\_0\_1\_2} and assume $h=0.73$.\\
Column 1: galaxy identification number assigned in \cite{2020ApJ...888...36R}. Column 2: galaxy name. Column 3: NSA identification number. Columns 4 \& 5: right ascension, declination, respectively, of the VLA radio detections from \citet[][]{2020ApJ...888...36R}. Column 6: redshift. Column 7: log galaxy stellar mass in units of $M_{\odot}$. Column 8: absolute $g$-band magnitude. Column 9: $g$$-$$r$ color. Column 10: Petrosian 50\% light radius. Column 11: Sérsic index, $n$.}
\label{tab:radio_obs}
\end{deluxetable*}

\begin{table*}
\caption{Observation Parameters}
\begin{center}
\begin{tabular}{ccccrcll}
\hline
\hline
\multicolumn{8}{c}{All Observations}\\
\hline
\hline
\multicolumn{4}{l}{Parameter}                           & \multicolumn{4}{l}{Value} \\
\hline
\multicolumn{4}{l}{Backend System}                      & \multicolumn{4}{l}{ROACH Digital Backend (RDBE)} \\
\multicolumn{4}{l}{Total IF windows}                    & \multicolumn{4}{l}{4} \\
\multicolumn{4}{l}{X-band channel frequencies (MHz)}    & \multicolumn{4}{l}{8412.0, 8540.0, 8668.0, 8796.0} \\
\multicolumn{4}{l}{Single window bandwidth (MHz)}       & \multicolumn{4}{l}{128} \\
\multicolumn{4}{l}{No. of spectral channels per window} & \multicolumn{4}{l}{512} \\
\multicolumn{4}{l}{Total bandwidth at X-band (MHz)}     & \multicolumn{4}{l}{512} \\
\multicolumn{4}{l}{Frequency resolution (MHz)}          & \multicolumn{4}{l}{0.25} \\
\multicolumn{4}{l}{Polarization}                        & \multicolumn{4}{l}{Right-hand circular} \\
\multicolumn{4}{l}{Data rate (Gbps)}                    & \multicolumn{4}{l}{2} \\
\multicolumn{4}{l}{Sampling rate (bits)}                & \multicolumn{4}{l}{2} \\
\multicolumn{4}{l}{Observed central frequency (GHz)}    & \multicolumn{4}{l}{8.67} \\
\hline
\hline
\multicolumn{8}{c}{Individual Observations} \\
\hline
\hline
ID  & $T_{\mathrm{int}}$ & \multicolumn{3}{c}{Restoring Beam}               & Calibrator & R.A.             & Dec. \\\cline{3-5}
    & (s)                & ($\alpha\times\delta$; mas) & ($\alpha\times\delta$; pc)& (P.A.; deg) & IERS Name  & (h m s $\pm~\mu$s) & ($^\circ~'~''~\pm~\mu''$)  \\
\hline
2  & 3516 & $3.01 \times 1.11$ & $2.16 \times 0.79$ & $-$10.06 & 0007+171   &  00 10 33.99062310 $\pm$ 4.18   & {+}17 24 18.7612955 $\pm$ 71.4   \\
6  & 3488 & $3.70 \times 1.25$ & $1.30 \times 0.44$ & $-$18.70 & 0106+013   &  01 08 38.77110544 $\pm$ 2.05   & {+}01 35 00.3171844 $\pm$ 31.5   \\
25 & 3492 & $1.96 \times 1.25$ & $1.08 \times 0.69$ & $-$10.75 & 0902+490   &  09 05 27.46386114 $\pm$ 11.96  & {+}48 50 49.9653540 $\pm$ 157.7  \\
26 & 3491 & $1.90 \times 1.30$ & $1.76 \times 1.21$ & $-$15.78 & 0850+581   &  08 54 41.99641612 $\pm$ 6.54   & {+}57 57 29.9391251 $\pm$ 60.5   \\
28 & 3500 & $1.91 \times 1.27$ & $1.15 \times 0.77$ & $-$20.01 & 0850+581   &  08 54 41.99641612 $\pm$ 6.54   & {+}57 57 29.9391251 $\pm$ 60.5   \\
33 & 3492 & $3.75 \times 1.84$ & $3.65 \times 1.79$ & $-$4.06  & 0923+575   &  09 27 06.05345129 $\pm$ 21.62  & {+}57 17 45.3431338 $\pm$ 223.7  \\
48 & 3496 & $3.42 \times 1.98$ & $1.41 \times 0.82$ & $-$19.40 & 1025+031   &  10 28 20.40128976 $\pm$ 6.71   & {+}02 55 22.4719033 $\pm$ 213.0  \\
64 & 3492 & $2.50 \times 1.76$ & $1.71 \times 1.20$ & { }2.67  & 1137+123   &  11 40 27.74465895 $\pm$ 22.82  & {+}12 03 08.2705140 $\pm$ 807.4  \\
65 & 3496 & $2.34 \times 1.26$ & $1.48 \times 0.80$ & $-$13.20 & 1123+264   &  11 25 53.71191677 $\pm$ 2.34   & {+}26 10 19.9787016 $\pm$ 31.5   \\
77 & 3492 & $4.44 \times 1.13$ & $2.31 \times 0.59$ & $-$17.01 & 1200$-$051 &  12 02 34.22488808 $\pm$ 5.36   & $-$05 28 02.4909262 $\pm$ 160.8  \\
82 & 3500 & $2.21 \times 1.21$ & $1.20 \times 0.66$ & $-$13.36 & 1215+303   &  12 17 52.08196690 $\pm$ 3.11   & {+}30 07 00.6358732 $\pm$ 43.2   \\
83 & 3508 & $2.98 \times 1.27$ & $1.46 \times 0.62$ & $-$8.89  & 1221+071   &  12 23 54.62431532 $\pm$ 7.55   & {+}06 50 02.5721111 $\pm$ 194.6  \\
92 & 3500 & $3.71 \times 1.40$ & $1.67 \times 0.63$ & $-$16.96 & 1253$-$055 &  12 56 11.16657958 $\pm$ 2.17   & $-$05 47 21.5251510 $\pm$ 38.8   \\
\hline
\end{tabular}
\end{center}
\tablecomments{The observing parameters for the targets. Phase calibrator and its position is the IERS name from the ICRF3 catalog.\\
Column 1: galaxy identification number. Column 2: observing time on target. Columns 3: restoring beam in mas. Column 4: restoring beam in pc (if associated with galaxy). Column 5: restoring beam position angle. Column 6: phase calibrator source. Column 7: right ascension of calibrator source. Column 8: declination of calibrator source.}
\label{tab:observations}
\end{table*}

\section{The Sample \label{sec:sample}}
To construct a sample of dwarf galaxies for observation with the VLA, \cite{2020ApJ...888...36R} drew from the NASA-Sloan Atlas (NSA; \texttt{v0\_1\_2}),\footnote{\url{http://www.nsatlas.org/}} which contains photometric and spectroscopic properties of SDSS galaxies with redshifts $z<0.055~(D\lesssim225~\mathrm{Mpc})$. The initial parent sample consisted of $43,707$ galaxies after selecting on the NSA for sources with stellar masses $M_{\star}\leqslant3\times10^9M_{\odot}$ and absolute magnitudes $M_g$ and $M_r>-20$. These cuts were chosen in part to help reduce luminous and massive galaxies with erroneous mass estimates, and with the mass limit approximately equal to the stellar mass of the LMC, the most massive dwarf satellite of the Galaxy. 

This set of galaxies was then cross-matched with the VLA Faint Images of the Radio Sky at Twenty-centimeters survey \citep[FIRST;][]{1995ApJ...450..559B}, allowing for a match tolerance of ${\leqslant}5\arcsec$ (approximately the angular resolution of FIRST), and found 186 matches. After interlopers were removed and scheduling priorities were considered, 111 of these 186 dwarf galaxies were followed up in the X-band (8$-$12 GHz) using the VLA in A~configuration ($0\farcs25$ angular resolution).

These observations detected compact radio objects toward 39 galaxies, four of which were determined to be likely background AGNs. The remaining 35 galaxies were split into two samples, one of which consisted of bona fide dwarf galaxies with reliable redshifts (Sample A: 28 galaxies) and the other with potentially more massive galaxies having less reliable redshifts (Sample B: 7 galaxies). After evaluating whether or not star formation-related processes, such as \ion{H}{2} regions or populations of supernovae/supernova remnants (SNe/SNRs), could plausibly account for the detected radio emission, \citet{2020ApJ...888...36R} determined that the radio sources towards 13 dwarf galaxies (in Sample A) were likely accreting MBHs. We selected these 13 dwarf galaxies for VLBA follow-up observations.
Table \ref{tab:radio_obs} lists the global properties from \cite{2020ApJ...888...36R} of the dwarf galaxies that we observed with the VLBA. 

\section{VLBA Observations \label{sec:observations}}
We obtained VLBA observations of the 13 sample galaxies in January 2020 as part of the United States Naval Observatory's 50\% VLBA timeshare allocation. Using the 9 GHz flux densities published in \cite{2020ApJ...888...36R}, we determined that an on-source integration time of 60 minutes per target was required in order to achieve a point-source detection threshold of $5\sigma$ using the online European VLBI Network Calculator.\footnote{http://old.evlbi.org/cgi-bin/EVNcalc.pl} We organized the 13 targets by right ascension and created three groups for which we formed three schedule files. Grouping the targets together in this way maximizes $uv$ coverage by alternating between sources in 20 minute intervals until a nominal observation of 60 minutes per source is reached. VLBA observations were centered on the positions of the VLA radio detections.  We used phase referencing by alternating four minute integrations of each target source with two minute integrations of a nearby calibrator source in order to obtain accurate phase calibrations over the span of the entire observation. All ten VLBA antennas participated in each of our three observing blocks for full angular resolution and maximum $uv$ sky coverage, but the Hancock antenna was removed during calibration for ID~48 due to strong RFI. The theoretical RMS noise for the VLBA observations is therefore 32 $\rm {\mu Jy~bm^{-1}}$ for all targets except for ID 48 which had a theoretical RMS noise of 36 $\rm {\mu Jy~bm^{-1}}$. The observation and instrumental setup parameters are noted in Table~\ref{tab:observations}. 

We calibrated our VLBA data using the National Radio Astronomy Observatory (NRAO) Astronomical Image Processing System \citep[\textsc{aips};][]{1996ASPC..101...37V}, release 31DEC20. We loaded the data using a calibration table interval of 0.1 minutes. Using the \textsc{vlbautil} module, we calibrated for the ionosphere and earth orientation parameters for phase-referenced observations, then for the correlator sampler threshold errors. Next, we calculated phase and delay calibrations, followed by bandpass, amplitude, and parallactic angle calibrations. We flagged the data for each baseline on both frequency and time versus amplitude for each source using \textsc{wiper}. We then fringe-fit the data for both the phase calibrator and the target to solve for complex amplitudes and phase where we used a solution interval of $0.25~\rm min$ for all targets except for ID~25 and ID~64, in which we used a solution interval of $0.5~\rm min$ and $2~\rm min$, respectively, to increase the number of good solutions when fringe-fitting on the calibrator source. Finally, we applied the calibrations to each source and split them out for imaging.  The VLBA flux density values are accurate to within the standard, nominal $5\%$ amplitude calibration uncertainty.

We imaged the calibrated data using the standard CLEAN algorithm in \textsc{aips} with the task \textsc{imagr}. Here, we used a pixel size of $0.5~\rm mas$, which Nyquist-samples the restoring beam, and a Briggs weighting scheme using \texttt{robust=5}, which is roughly equal to natural weighting.  We started with an image size of 512 pixels per side and increased this to 1024, 2048 and 4096 pixels for images that did not have any obvious detections.  However, the astrometric accuracy of the VLA data was within $0\farcs1$, so it is unlikely that the VLBA observations missed any point sources detected by the VLA. In this work, we define a detection using standard noise statistics with a signal-to-noise ($S/N$) threshold above the $5\sigma$ level to search for evidence of radio continuum structure ($\gtrsim$ the synthesized beam) using \textsc{imstat} within \textsc{aips}. Non-detections are defined from their white noise maps with no discernible emission above a $S/N$ of 5.

\begin{figure*}
\includegraphics[width=\textwidth]{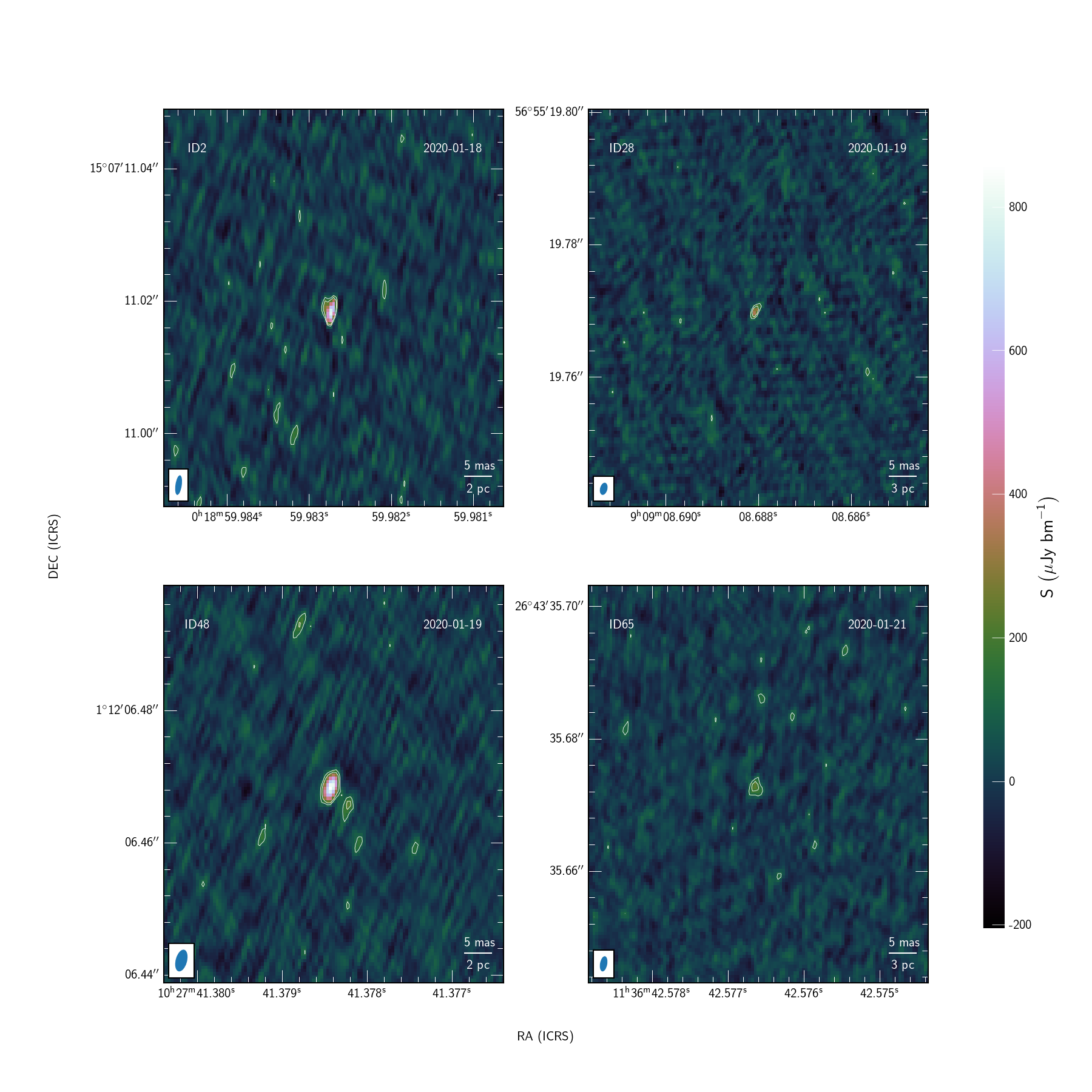}
\caption{The VLBA detected sources with contour lines depicting RMS levels from Column 2 of Table \ref{tab:vlba_detections_sensitivity} at the $3\sigma$ and $5\sigma$ levels. The physical scale assumes that the sources are at the same distance as their associated dwarf galaxies, but these sources are likely background AGNs (Section~\ref{subsec: nature of the detections}).}
\label{fig:detections}
\end{figure*}

\begin{deluxetable*}{CCCCCCCC}
\tablecaption{VLBA measurements for detections \label{tab:vlba_measurements}}
\centering
\tablehead{
\colhead{ID} & 
\colhead{R.A.} & 
\colhead{Dec.} & 
\colhead{$F_{\mathrm{peak}}$} & 
\colhead{$\log{L_{\mathrm{peak}}}$} & 
\colhead{$S_{\mathrm{tot}}$} &
\colhead{offset} &
\colhead{size} \\  
& 
\colhead{(h m s $\pm~\mu s$)} & 
\colhead{($^\circ~'~''\pm\mu''$)} & 
\colhead{($\mu$Jy ${\rm bm}^{-1}$)} & 
\colhead{($\rm erg~s^{-1}~Hz^{-1}$)} & 
\colhead{($\mu$Jy)} &
\colhead{($''$)} &
\colhead{(pc)}
}
\startdata
2  & 00~18~59.985 \pm ~4.59 & +15~07~11.02 \pm ~89.9 & 831 \pm 57 & 28.40 & 874 \pm 83  & $4.9$ & ${<}1.95 \pm 0.14$ \\
28 & 09~09~08.689 \pm 10.66 & +56~55~19.75 \pm 112.8 & 354 \pm 44 & 27.87 & 323 \pm 67  & $2.7$ & ${<}1.33 \pm 0.17$ \\
48 & 10~27~41.380 \pm ~7.16 & +01~12~06.45 \pm 220.4 & 899 \pm 59 & 27.93 & 790 \pm 73  & $4.2$ & ${<}1.32 \pm 0.09$ \\
65 & 11~36~42.578 \pm ~9.64 & +26~43~35.66 \pm 140.5 & 245 \pm 37 & 27.75 & 375 \pm 84  & $2.9$ & ${<}1.43 \pm 0.22$
\enddata
\tablecomments{Column 1: galaxy identification number. Column 2: measured right ascension. Columns 3: measured declination. Column 4: peak flux density with uncertainties listed in Column 2 of Table \ref{tab:vlba_detections_sensitivity} and including an additional 5\% systematic uncertainty.  Column 5: log peak luminosity (if associated with galaxy). Column 6: integrated flux density. Column 7: detection offset from photo-center as depicted in the NSA (\texttt{v0\_1\_2}). Column 8: major axis diameter (if associated with galaxy), expressed as an upper limit because the VLBA detections are unresolved.}
\end{deluxetable*}

\begin{deluxetable*}{lCCCCCCCC}
\tablecaption{Sensitivity - Detections \label{tab:vlba_detections_sensitivity}}
\tablehead{
\colhead{ID} & 
\sigma_{\rm VLBA} & 
\sigma_{\rm VLA} & 
S_{\mathrm{tot,VLA}} & 
\log L_{\rm{VLA}} &
T_{\mathrm{B,VLBA}} &
S_{\mathrm{tot}}/F_{\mathrm{peak}} & 
S/N & 
S_{\mathrm{tot}}/S_{\mathrm{tot,VLA}} \\
&
\multicolumn2c{($\rm \mu Jy~bm^{-1}$)} & 
(\rm \mu Jy) & 
&
(\rm erg~s^{-1}~Hz^{-1}) &
($\times10^4~\mathrm K$) &
&
}
\startdata
2  & 40 & 14 & 2586 $\pm$ 132 & 28.9 & {>}404.34 & 1.05 $\pm$ 0.12 & 20.74 & 0.34 $\pm$ 0.11 \\
28 & 40 & 13 & 595  $\pm$ 37  & 28.1 & {>}237.25 & 0.91 $\pm$ 0.24 & ~8.89 & 0.54 $\pm$ 0.22 \\
48 & 39 & 17 & 2587 $\pm$ 133 & 28.4 & {>}215.82 & 0.88 $\pm$ 0.11 & 22.95 & 0.31 $\pm$ 0.11 \\
65 & 36 & 13 & 517  $\pm$ 35  & 28.1 & {>}135.09 & 1.53 $\pm$ 0.27 & ~6.81 & 0.73 $\pm$ 0.24
\enddata
\tablecomments{Measurements for the detections in the VLBA observations.
Column 1: galaxy identification number. Column 2: measured RMS for the VLBA observations. Column 3: measured RMS noise for the VLA observations presented in \citet[][]{2020ApJ...888...36R}. Column 4: VLA flux densities at 9 GHz, with uncertainties listed in Column 3 and including an additional 5\% systematic uncertainty.  Column 5: spectral luminosities from the VLA observations, in erg~s$^{-1}$~Hz$^{-1}$. Column 6: lower limits on the brightness temperatures for the VLBA detections. Column 7: ratio of integrated flux density to peak flux density for the VLBA detections. Column 8. signal to noise ratio for the VLBA detections. Column 9. VLBA to VLA integrated flux density ratios.}

\end{deluxetable*}

\begin{deluxetable*}{lCCCCCC}
\tablecaption{Sensitivity - Non-detections \label{tab:vlba_nondetections_sens}}
\tablehead{
\colhead{ID} & 
\sigma_{\rm VLBA} & 
\sigma_{\rm VLA} & 
S_{\mathrm{VLA}} & 
3\sigma_{\rm VLBA}/L_{\rm VLA}& 
T_{\rm VLA}<T_{\rm B}<T_{\rm VLBA} & 
S_{\mathrm{VLA}}/\sigma_{\mathrm{VLBA}}\\
& 
\multicolumn2c{($\rm \mu Jy~bm^{-1}$)} & 
(\rm \mu Jy) & 
& 
(\times 10^4~\rm K)
}
\startdata
6  & 42 & 14  & 352  \pm 30              & {<}0.41 & 0.01 < T_{\rm B} < 0.59 & ~~8 \\
25  & 45 & 15 & 375  \pm 47              & {<}0.44 & 0.02 < T_{\rm B} < 0.63 & ~~8 \\
26$^{\dagger}$  & 41 & 13 & 929  \pm 53  & {<}0.15 & 0.04 < T_{\rm B} < 1.56 & ~22 \\
33 & 59 & 13 & 1951 \pm 101              & {<}0.09 & 0.09 < T_{\rm B} < 3.27 & ~33 \\
64  & 21 & 13 & 3218 \pm 163             & {<}0.02 & 0.17 < T_{\rm B} < 5.39 & 153 \\
77  & 47 & 17 & 1496 \pm 82              & {<}0.10 & 0.04 < T_{\rm B} < 2.51 & ~31 \\
82$^{\dagger}$ & 41 & 14 & 397  \pm 31   & {<}0.31 & 0.03 < T_{\rm B} < 1.31 & ~~9 \\
83  & 59 & 15 & 780  \pm 47              & {<}0.23 & 0.04 < T_{\rm B} < 1.94 & ~13 \\
92  & 42 & 24 & 1160 \pm 96              & {<}0.11 & 0.02 < T_{\rm B} < 0.67 & ~27 \\
\enddata
\tablecomments{Measurements for the non-detections in the VLBA observations.
Column 1: galaxy identification number. Column 2: measured RMS for the VLBA observations. Column 3: measured RMS noise for the VLA observations from \citet{2020ApJ...888...36R}. Column 4: VLA flux densities at 9 GHz from \citet{2020ApJ...888...36R}, including 5\% systematic uncertainties. Column 5: VLBA radio luminosity upper limits as a fractions of VLA luminosities at the 3$\sigma$ level. Column 6: brightness temperature ranges corresponding to the VLA at its angular resolution limit and the VLBA at its largest angular scale. Column 7: upper limit for non-detections at the $1\sigma$ level.\\\\
${\dagger}$ ID~26 and ID~82 show optical spectroscopic signatures consistent with accreting MBH  \citep{2013ApJ...775..116R, 2021ApJ...910....5M}, and ID~26 also shows X-ray signatures consistent with accreting MBH \citep[ID~9 in][]{2017ApJ...836...20B}.
}
\end{deluxetable*}

\section{Results \label{sec:results}}
We detect a total of 4 radio sources out of our 13 targets, shown in Figure~\ref{fig:detections}. In Table \ref{tab:vlba_measurements} we give the measurements for the detected sources and find that the detections are unresolved point sources, as the integrated flux density versus peak flux density ratios are on the order of unity. The VLBA-detected source measurements in Table~\ref{tab:vlba_measurements} were calculated using \textsc{jmfit} to fit a single-component 2-D Gaussian model to the detection to measure right ascension and declination, as well as a peak flux density and integrated flux density. Assuming that they are truly associated with their dwarf galaxies (but see Section~\ref{subsec: nature of the detections}), their luminosities are on the order of $10^{27}-10^{28}$ $\mathrm{erg~s^{-1}~Hz^{-1}}$, and their physical extent is ${<}2$~pc.

In Tables \ref{tab:vlba_detections_sensitivity} and \ref{tab:vlba_nondetections_sens} we list the sensitivities of all the observations. We measured the thermal noise of our observations by selecting a region that did not contain any detected emission. We used the final, cleaned images to measure the RMS for the VLBA detections and a lightly cleaned image with 200 iterations for the non-detections to deconvolve the point spread function. The measured thermal noise for both VLA and the VLBA observations are listed in columns 2 and 3 in Tables \ref{tab:vlba_detections_sensitivity} and \ref{tab:vlba_nondetections_sens}.  Our measured RMS is within a factor of ${<}2$ of the theoretical value, derived using Equation~9-23 from \citet{1999ASPC..180..171W}\footnote{\url{https://science.nrao.edu/facilities/vlba/docs/manuals/oss2022A/imag-sens}} for a typical system equivalent flux density of $\rm{327~Jy}$ at $\rm{4~cm}$ for the VLBA\footnote{https://science.nrao.edu/facilities/vlba/docs/manuals/oss/bands-perf\#Table\%205.1} and an efficiency of 0.8\footnote{https://science.nrao.edu/facilities/vlba/docs/manuals/oss/bsln-sens}, so we achieved the designed observation sensitivities for all of our targets.

When we compare the integrated flux density between the VLA and VLBA for the detected sources, we find that the VLA detections as measured in \cite{2020ApJ...888...36R} are ${\sim}1-3$ times brighter than the measured flux densities when compared with our VLBA detections. 
We also calculate brightness temperatures for the VLBA detections and find that they fall in the range of $10^{6.1}$ $\mathrm{K}$ to $10^{6.6}$ $\mathrm{K}$, where the brightness temperature is calculated as:

\begin{equation}\label{eq:Tb}
T_b=\frac{4\ln(2)c^2}{2\pi k_B\nu^2\theta_{\mathrm{min}}\theta_{\mathrm{max}}}I_{\nu}
\end{equation}

\noindent Here, $I_{\nu}$ is the intensity in $\mathrm{Jy~bm^{-1}}$ (and is the measured peak flux density, $F_{\mathrm{peak}}$, for our detected point sources), $\nu$ is the frequency in $\mathrm{GHz}$, and $\theta_{\mathrm{max}}$ and $\theta_{\mathrm{min}}$ are the half-power beam widths along the major and minor axes in arcseconds. Given that the VLBA detections are point sources, the measured brightness temperatures are therefore lower limits.  We present the calculated limits of the brightness temperatures for the VLBA non-detections in Table \ref{tab:vlba_nondetections_sens} and discuss the implications of these in the following section.

\section{Discussion \label{sec:discussion}}
\subsection{The nature of the VLBA detections} \label{subsec: nature of the detections}
To determine the nature of the four VLA sources detected with the VLBA, we take into consideration their brightness temperatures, apparent luminosities, and location with respect to their associated optical dwarf galaxy stellar counterparts. The flux densities and lack of source extent indicates that all of the VLBA detections have brightness temperatures greater than ${>}10^6$~K, which strongly favors non-thermal emission from compact objects such as AGNs. To explore whether these sources could be powered by X-ray binaries (XRBs) or ultraluminous X-ray sources (ULXs), we checked the database of \citet{arash_bahramian_2018_1252036}, which contains radio and X-ray observation data for XRBs within our Galaxy. This database shows that the most luminous radio X-ray binaries in outburst are at or below $L \lesssim 10^{31}~{\rm erg~s^{-1}}$ at 5 GHz in the radio, or $L \lesssim 10^{21}~{\rm erg~s^{-1}~Hz^{-1}}$. By contrast, assuming that the four VLBA sources are at the same distance as their associated dwarf galaxies, the measured VLBA luminosities we have detected are around $L \approx 10^{28}~{\rm erg~s^{-1}~Hz^{-1}}$, seven orders of magnitude greater than what is observed in our Galaxy. It therefore seems unlikely that the four VLBA detections are X-ray binaries.

Compact point-like source emission has been detected in a number of nearby star forming dwarf irregular galaxies \citep[see e.g.,][]{Hindson_2018}.  For example, \cite{Cseh_2012} looked at ULXs in the radio, optical and X-ray associated with large scale nebulae by observing IC~342~X-1 and the dwarf irregular galaxies Holmberg~II~X-1 with the VLA and NGC~5408 X-1 with the Australia Telescope Compact Array (ATCA) and Very Large Telescope (VLT). They observed Holmberg~II~X-1 ($D=3.39$~Mpc) with the VLA at 8.5~GHz in the C-configuration and measured a flux density of $395 \pm 40~{\rm \mu Jy}$, corresponding to a spectral luminosity density of $L_\mathrm{8.5~GHz} = (5.43\pm 0.55) \times 10^{24}~{\rm erg~s^{-1}~Hz^{-1}}$. We find that the radio spectral luminosity densities calculated using their observations of NGC~5408~X-1 and IC~342~X-1 are $L_\mathrm{9~GHz} = (3.78 \pm 0.99) \times 10^{24}~{\rm erg~s^{-1}~Hz^{-1}}$ and $L_\mathrm{4.8~GHz} = (3.64 \pm 0.18) \times 10^{25}~{\rm erg~s^{-1}~Hz^{-1}}$, respectively; all orders of magnitude lower than our measured VLBA luminosities.

\cite{2020ApJ...888...36R} investigate the luminosities of 102 younger radio SNe in the merging infrared galaxies Arp 220 and Arp 299 by exploring the works by \cite{Ulvestad_2009} and \cite{Varenius_2019}, and they find that the vast majority of these sources have a spectral luminosity density of $L_{\nu}\lesssim 10^{27}~\rm erg~s^{-1}~Hz^{-1}$. The apparent luminosities of our VLBA detections are an order of magnitude larger, disfavoring interpretation as single SNe, and requiring ${\sim}10$ SNe within a ${\sim}1-2$~pc diameter volume.

\begin{figure}
\includegraphics[width=\columnwidth]{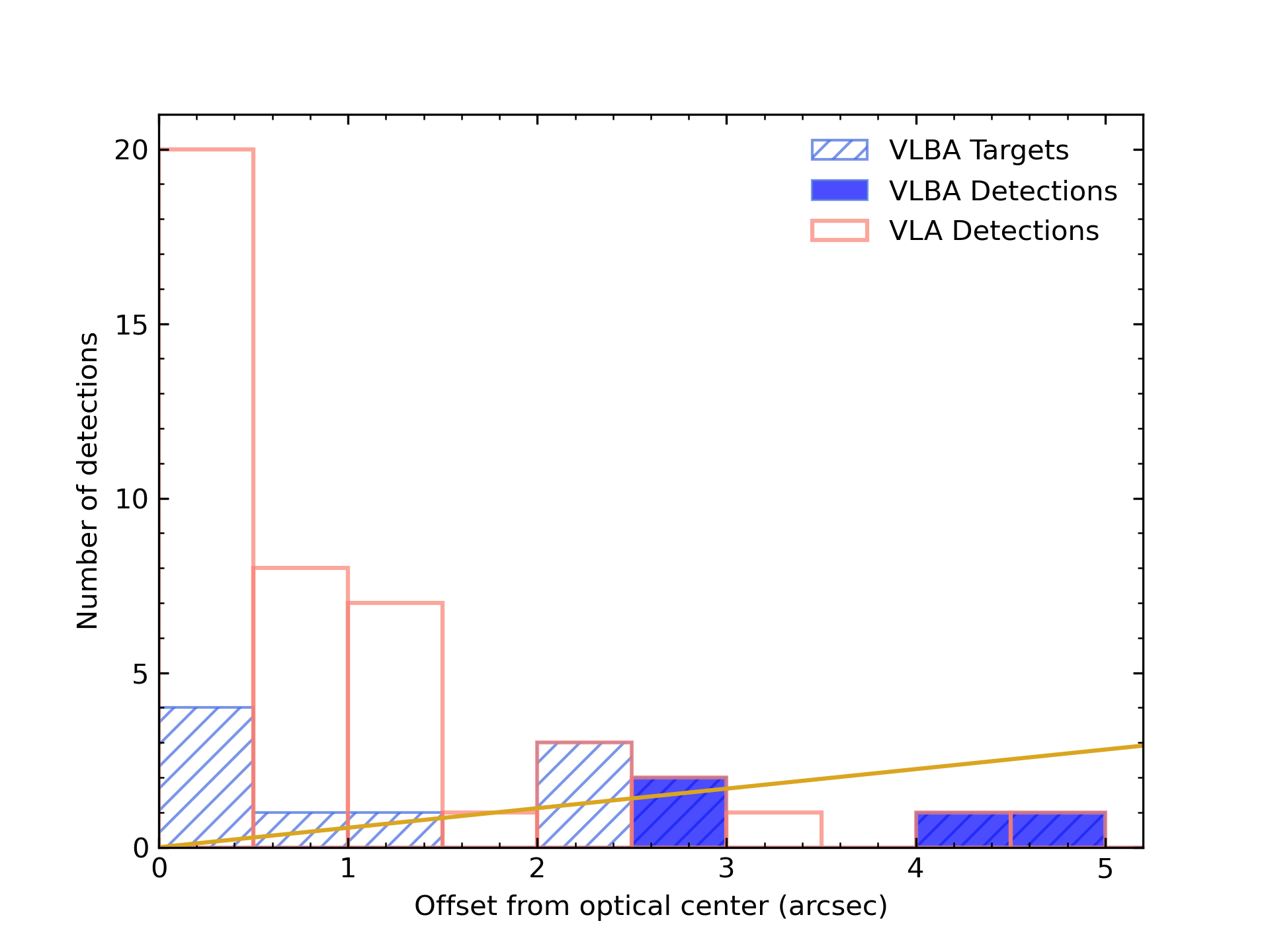}
\caption{Distribution of galaxy photo-center offsets for the 13 VLA sources targeted with the VLBA, as well as the other radio sources from \citet{2020ApJ...888...36R} detected with the VLA for reference. Photo-centers are constructed from Table 1 in \citet{2020ApJ...888...36R} and were originally determined from the NSA (\texttt{v0\_1\_2}). The yellow line shows the expected distribution for background AGNs, as calculated in \citet{2020ApJ...888...36R}. Notably, the four VLA sources detected by the VLBA are all at distances consistent with expectations for background AGNs, and are the furthest from the optical center of the 13 galaxies observed.} 
\label{fig:offsets}
\end{figure}

With these considerations, accretion onto MBHs is the more likely interpretation of the nature of the radio emission for these four VLBA detections, potentially in line with expectations for wandering MBHs. However, it is notable that these four VLBA detections are also the four furthest of the 13 VLA radio sources from the photo-centers of their associated dwarf galaxies. In Figure \ref{fig:offsets} we recreate the histogram shown in Figure 5 of \cite{2020ApJ...888...36R} and find that the four VLBA detections are offset from the galaxy photo-centers by $2\arcsec$ or more. Here, ID~28 and ID~65 are within $3\arcsec$ of the optical center and ID~2 and ID~48 are offset by more than $4\arcsec$. To test if these sources are consistent with background AGNs, we use the scaling relation from \citet[][their Figure~5, right panel]{2020ApJ...888...36R}, which gives the expected distribution of background radio sources such that $N(d_{\rm off})=A \times 1.6~d_{\rm off}$, where $A=0.35$ is a scaling factor for seven expected background sources within a $5\arcsec$ radius from the photo-center, and $d_{\rm off}$ is the offset in arcseconds. As can be seen in Figure \ref{fig:offsets}, the four VLBA detections are at distances consistent with expectations for background AGNs. 

Moreover, as noted above, the four VLBA detections are the most distant of the 13 sources from their associated galaxies. Randomly assigning some property (i.e., VLBA detections) to four objects from a sortable list of size 13, the probability of getting this result is $p=0.0014$. This strongly implies that there is a preference for objects being detectable with the VLBA at larger distances from the host galaxy, which is in agreement with their interpretation as background AGNs. Given their photo-center distances, and the statistical considerations outlined here, the four VLBA detections are likely background AGNs.


\subsection{The nature of the VLBA non-detections}
In Table \ref{tab:vlba_nondetections_sens}, we provide the 
expected VLBA signal-to-noise ($S/N$) assuming that the entire VLA flux density is contained in a compact core on milliarcsecond scales. If these objects are indeed point sources, then we should have detected all of them at a $S/N \ge 5$, well above a reasonable noise floor. The lack of detections with the VLBA indicates that these sources have likely been ``resolved out'' due to the dramatically increased angular resolution of the VLBA and lack of short baselines, so therefore exhibit source extent beyond the largest angular scale (LAS) of our $8.67~\rm GHz$ VLBA observations: ${\sim}30$ mas, or about 16~pc at the ${\sim}110$~Mpc median distance of the non-detections.

Compact AGN cores could still be present if they have a peak brightness well below the flux density measured with the VLA on larger scales. This type of partial resolution of AGN emission is common. For example, a VLBI survey of ${\sim}25,000$ FIRST sources that are expected to be dominated by radio AGN found that only 30-40\% had a milliarcsecond-scale core with a peak brightness ${\ge}32$\% of the flux density detected with the VLA on arcsecond scales \citep{2014AJ....147...14D}. In other words, while detection with the VLBA supports the presence of AGNs, non-detections do not necessarily rule them out. 
For example, both of the dwarf starburst galaxies Henize 2-10 \citep[e.g.,][]{schuttereines2022} and Mrk 709 \citep{Kimbro2021} were detected in VLA observations \citep[][]{reines2011,reines2014}, yet neither were detected with the VLBA \citep[][]{ulvestad2007,Kimbro2021}. On the other hand, the median distance to the non-detections is 12~times further than Henize~2-10, so the VLBA beam subtends dramatically different physical scales. Indeed, core emission was detected in Henize~2-10 using Long~Baseline~Array (LBA) observations \citep{reinesdeller2012} that, with an angular resolution of $0\farcs1\times0\farcs03$, subtend nearly the same physical scale as our VLBA observations, ${\sim}1-3$~pc. The radio emission in the VLBA non-detections is therefore likely to be more extended than the source in Henize~2-10. However, while the LBA radio core of Henize~2-10 was found to have a brightness temperature (Equation \ref{eq:Tb}) of at least $>3\times10^5$~K \citep{reinesdeller2012}, closer to expectations for compact non-thermal emission driven by a MBH, it also has a spectral luminosity of $\sim10^{26}$~erg~s$^{-1}$~Hz$^{-1}$, which would be undetectable by both the VLBA and VLA observations at the median distance of our sample.

We note that, with the exception of IDs 25 and 92, the non-detections were found by \citet{2020ApJ...888...36R} to be unresolved with the VLA, setting a lower bound on the brightness temperatures of a few times $\times10^2$~K. These brightness temperatures provide a range of possible values that are consistent with star formation-related processes. However, such brightness temperatures are not inconsistent with extended radio emission seen in the vicinity of AGNs due to winds or jet lobes. We note that the VLBA non-detections include ID~26 and 82, which were confirmed to host AGNs \citep{2013ApJ...775..116R,2017ApJ...836...20B,2021ApJ...910....5M}. ID~26 has Seyfert-like narrow emission line ratios and exhibits broad H$\alpha$ emission consistent with a MBH with $M_{\rm BH} \sim 2.5 \times 10^5~M_\odot$ (ID~9 in \citealt{2013ApJ...775..116R}). ID~82 exhibits the AGN coronal line [\ion{Fe}{10}], enhanced [\ion{O}{1}] emission coincident with the radio source, and broad H$\alpha$ emission \citep{2021ApJ...910....5M}. On the other hand, the VLA radio positions of ID~26 and 82 are within $0\farcs4$ of their measured host galaxy photo-centers and, as far as we can tell, are nuclear. Four of the VLBA non-detections, specifically IDs 26, 82, 83, and 92, have VLA source photo-center offsets less than $0\farcs5$, where the uncertainty of the photo-center position is an important consideration in deciding if an AGN is off-nuclear. While the NSA catalog does not give position uncertainties, comparison with the SDSS positions shows that 90\% of the 9 VLBA non-detections have NSA and SDSS positions that agree within $\sim0\farcs6$, giving some sense of the photo-center position uncertainties. It is therefore possible that the VLA sources in these four objects are not off-nuclear. Finally, star formation is often concentrated at the centers of AGN host galaxies \citep[e.g.,][]{2014ApJ...780...86E}, so co-spatial radio emission is not necessarily related to AGN activity.

Interpretation of the origin of the VLA radio emission in the 9 VLBA non-detections depends on the likelihood of their emission being star formation-related, versus either wandering MBHs or background AGNs. \citet{2020ApJ...888...36R} presented evidence disfavoring the star formation interpretation for the 13 galaxies observed in our program. First, they consider \ion{H}{2} regions and calculate the radio-based star formation rate (SFR) of a VLA source under this assumption. These are compared to galaxy-wide SFRs calculated using a combination of GALEX FUV and WISE 22~\micron\ data, although the latter was not available for nine of the 13 galaxies, four of which are in our sample of VLBA non-detections. If the implied SFR of the radio source (assuming it is a thermal \ion{H}{2} region) is significantly above the galaxy-wide SFR, the VLA source likely has a different, non-\ion{H}{2} region origin (e.g., SNe/SNRs or an AGN) since a single star-forming region cannot have a SFR greater than its entire host galaxy. The candidate MBHs that are the focus of this work were found to be significantly more luminous in the radio than the expectation from their SFRs. While the FUV+IR-based relationship used by \citet{2020ApJ...888...36R} gives SFRs consistent with H$\alpha$-based SFRs to within $\pm0.13$~dex \citep{2011ApJ...741..124H}, additional systematic uncertainties may affect the accuracy of the implied SFRs.

\begin{figure*}
  \centering
  \begin{subfigure}
    \centering
    \includegraphics[width=0.45\textwidth]{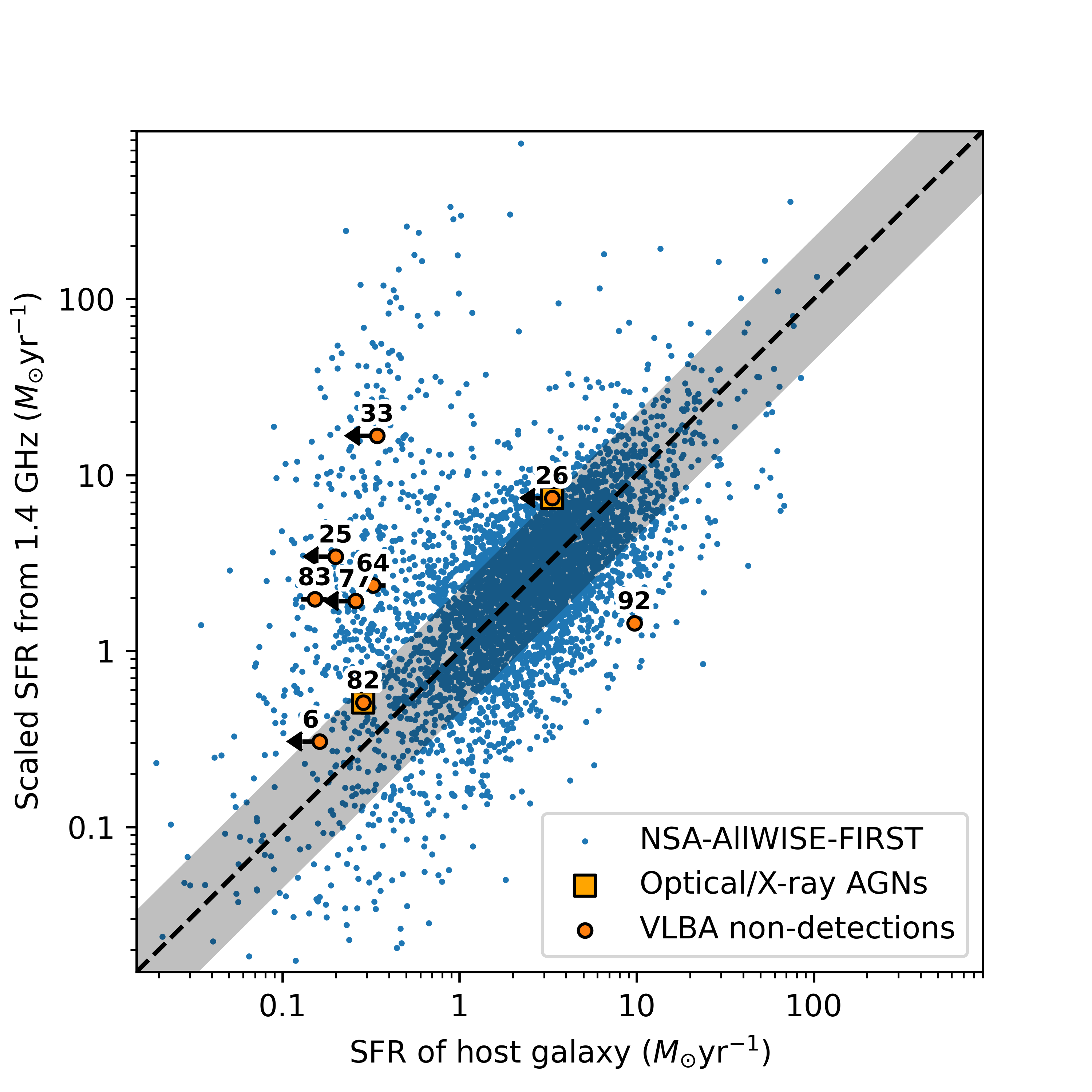}
    \label{fig:sfr_L_vs_fuv}
  \end{subfigure}%
  \quad
  \begin{subfigure}
    \centering
    \includegraphics[width=0.45\textwidth]{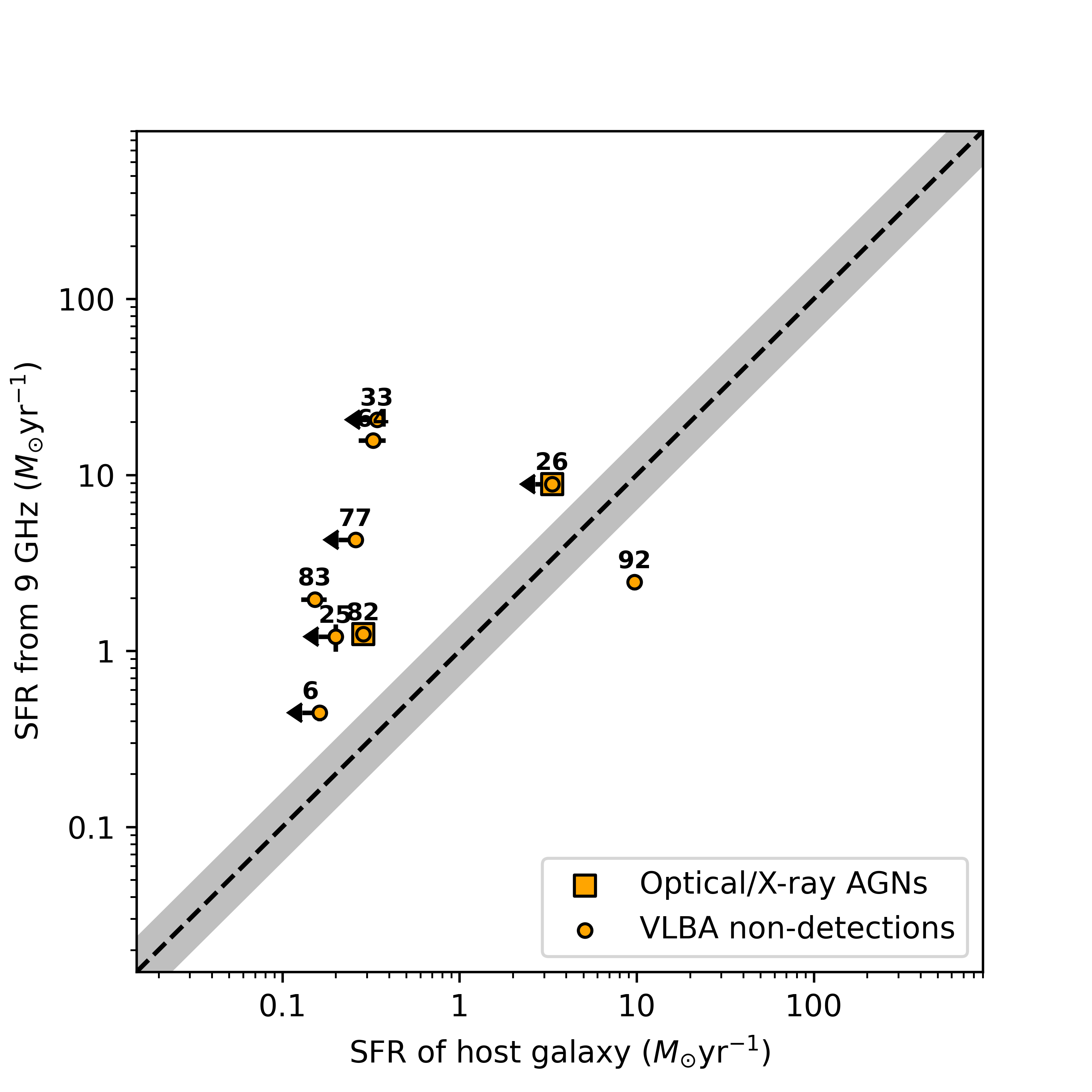}
    \label{fig:sfr_X_vs_fuv}
  \end{subfigure}
  \caption{Left: SFRs from 1.4~GHz FIRST data for the full NSA catalog and VLBA non-detections, compared to FUV-based SFRs, after correction for a 1.3~dex offset consistent with expectations for the ratio between the total and thermal 1.4~GHz luminosities for star-forming galaxies. The gray region denotes a dispersion of 0.35~dex, which we estimated directly from the data. Right: SFRs calculated directly from the \citet{2020ApJ...888...36R} X-band luminosities, using the total radio luminosity calibration of \citet{2011ApJ...737...67M}, with a 0.2~dex dispersion estimated from the comparison of different SFR indicators in \citet{2011ApJ...737...67M}.}
  \label{fig:sfrs}
\end{figure*}

To explore this last point, we have extended the SFR analysis as done in \citet[]{2020ApJ...888...36R}, performing two tests. First, we repeated the calculation from \citet[][Section~5.1]{2020ApJ...888...36R} in which all of the radio emission is assumed to be thermal and the corresponding star formation rate is calculated given the relationship between thermal radio luminosity and Lyman continuum photons from \citet{1992ARA&A..30..575C}. This is checked against the star formation rate calculated using the reddening-corrected far-UV (FUV) luminosity, which uses the WISE~W4 [$22\micron$] luminosity for the correction. This time, however, we used the entire NSA~catalog, not just the target dwarf galaxies, by cross-matching it with AllWISE and FIRST and used the FIRST flux densities for the star formation rate calculation due to radio emission. For the full NSA catalog, we required all objects to have both FUV and W4 measurements, but for the non-detections we allowed objects with ``reconstructed'' FUV magnitudes from the NSA spectral energy distribution fits, as well as objects with only 95\% confidence upper limits on their W4 mag. For these cases, we treat the FUV-based SFRs as upper limits. The radio continuum below $\sim30$~GHz is synchrotron-dominated, so as expected we find a large offset, by $\sim1.3$~dex, between the SFR calculated using the 1.4~GHz luminosities and the FUV$+W4$-based SFRs. Indeed, using the radio spectral energy distribution of the prototypical starburst galaxy M82 \citep[e.g.,][Figure~1]{1992ARA&A..30..575C}, the ratio between the total and thermal 1.4~GHz luminosity is $\sim1.2$~dex, almost exactly the overall offset that we find. We therefore correct the radio-based SFRs using the 1.3~dex offset, and compare the non-detections in this study to the overall galaxy population. We find a dispersion between 1.4~GHz- and FUV-based SFRs of $\sim0.35$~dex (Figure~\ref{fig:sfrs}, left), which places IDs 6, 26, 92, and 82 firmly within statistical expectations for star formation-powered radio emission. IDs 64 and 77 are possibly also consistent, although IDs~64 and 77 are near the base of a plume of objects with high radio-based SFRs that are likely radio AGNs. We note that ID~26 exhibits AGN-like mid-IR colors \citep{2021ApJ...914..133L}, and indeed ID~26 is a spectroscopic/X-ray AGN \citep{2013ApJ...775..116R, 2017ApJ...836...20B}, so its $W4$~[$22\micron$] emission may be contaminated. We therefore treat its SFR as an upper limit in Figure~\ref{fig:sfrs}. Given these considerations, about $\sim4$ of the 9 VLBA non-detections exhibit radio emission consistent with star formation-powered emission in the broader galaxy population.

For the second test, we directly calculated the SFRs expected from the total X-band luminosities found in \cite{2020ApJ...888...36R}. We used the \citet{2011ApJ...737...67M} calibration, which gives the total thermal~$+$~non-thermal radio emission at a particular frequency, given an electron temperature $T_e$ and non-thermal spectral index $\alpha^{\rm{NT}}$. Adopting $T_e=10^4~\rm K$ and a typical synchrotron spectral index of $\alpha^{\rm{NT}}=0.75$, we find results very similar to our first analysis (Figure~\ref{fig:sfrs}, right), with IDs 6, 26, and 92 consistent with expectations for star formation. ID~82 is less consistent with star formation than before, although in this case the statistical dispersion is taken at face value from the SFR indicator analyses in \citet{2011ApJ...737...67M} because we do not have comparable X-band data for the full NSA catalog. We adopt a value of 0.2~dex, which is based on the comparison of different SFRs performed in \citet{2011ApJ...737...67M} for all but the 1.4~GHz luminosity-based SFR estimator, which has a scatter of 0.35~dex. From this analysis, IDs~25, 33, 64, 77, possibly 82, and 83 appear to be too radio luminous for their star formation rates. We note that in the first test the radio emission from FIRST is well-matched to the angular resolution of the GALEX FUV data ($\sim5\arcsec$), while the X-band sources from \cite{2020ApJ...888...36R} are much more compact, subtending $\lesssim0\farcs2$. This raises the question of whether or not more extended radio emission seen in FIRST is being resolved out in the X-band data, the latter of which was taken in VLA A~configuration. We view this as being unlikely, for the following reasons. First, the largest angular scale of the VLA in A~configuration at X-band is $5\farcs3$,\footnote{\url{https://science.nrao.edu/facilities/vla/docs/manuals/oss/performance/resolution}} the same as the angular resolution of the FIRST data. Second, the ratio between the integrated and peak FIRST flux densities is close to unity for all of the non-detections, indicating that it is unlikely that there is radio emission from scales outside of detectability in the X-band data. Finally, the X-band observations achieved about an order of magnitude greater depth than the FIRST data, so even steep-spectrum emission present in the 1.4~GHz FIRST data should have been detected in the 9~GHz X-band data. With these considerations, we do not consider differences in angular scale to be significant for our second test.

An additional potential origin for non-AGN radio emission in the non-detections are individual, luminous SNRs. To test this, \citet{2020ApJ...888...36R} compare the luminosities of the VLA X-band sources to expectations for the most luminous SNR in a given galaxy, determined from the work of \citet{2009ApJ...703..370C}, who study 19 nearby star-forming galaxies to derive the SNR luminosity function. All of the radio sources in question are much too luminous to be produced by single SNRs, although \citet{2009ApJ...703..370C} note that SNRs in M82 and NGC~253 are as luminous as ${\sim}10^{27}$~erg~s$^{-1}$~Hz$^{-1}$ at 1.4~GHz, similar to the 1.4~GHz luminosity of ID~6 (${\sim}8\times10^{27}$~erg~cm$^{-2}$~s$^{-1}$). The normalization of the SNR luminosity function scales with SFR, however, and the SNR-driven radio luminosities of the non-detections are much larger than expected given their SFRs. On the other hand, the prevalence of luminous SNe and SNRs is fundamentally stochastic, so SNe/SNRs may exist that are significantly more luminous than expected for their host galaxy SFR \citep[as indeed is the case for the SNe-powered superbubble in IC~10;][]{2009ApJ...703..370C}. \citet{2009ApJ...703..370C} estimate the statistical sampling error of the most luminous SNRs as a function of SFR (their equations 25 and 26), which \citet{2020ApJ...888...36R} use to demonstrate that the objects considered here are significantly more luminous than expected. Finally, \citet{2020ApJ...888...36R} compare the observed compact radio luminosities to the expected cumulative luminosity from a population of SNRs/SNe in the entire host galaxy, making use of the radio SNR luminosity function from \citet{2009ApJ...703..370C}, again finding that the objects studied here are too luminous. It is important to note, however, that the number of galaxies considered in \citet{2009ApJ...703..370C} is small and selected from the local volume, and \citet{2009ApJ...703..370C} caution that their luminosity functions are poorly sampled at the most high-luminosity end. Indeed, the parent sample of \citet{2020ApJ...888...36R} are 186 dwarf galaxies with radio emission luminous enough to be detected by FIRST, selected from a much larger sample of several tens of thousands of dwarf galaxies not detected, so selection on objects with atypically bright SNRs or unusually prevalent SNe may be a factor. 

Finally, we reiterate that IDs 26 and 82, which have radio luminosities consistent with a star formation origin, were confirmed to host AGNs, so we emphasize that radio emission consistent with star formation does not preclude the existence of accreting MBHs in the other non-detections. Follow-up studies, such as with high angular resolution X-rays (e.g., with the Chandra X-ray Observatory), spectroscopic observations sensitive to high-ionization coronal lines, or multi-epoch radio observations to probe for variability that would indicate a compact emitter at a luminosity below the sensitivity of our VLBA observations, may alleviate the ambiguity between star forming processes and massive black holes.  We also reiterate that background AGNs may have a lingering presence in the non-detections, although the fact that the background AGNs discussed in Section~\ref{subsec: nature of the detections} were detected with the VLBA somewhat disfavors this scenario.

\section{Summary and Conclusions \label{sec:conclusion}}

We observed with the VLBA 13 radio-selected AGN candidates in dwarf galaxies from
\citet[][]{2020ApJ...888...36R}
as a follow-up study to potentially confirm the existence of wandering MBHs. Our VLBA observations have a beam solid angle that is smaller by a factor of ${\sim}10^4$ compared to the VLA observations of \citet[][]{2020ApJ...888...36R}, and comparable point-source sensitivity.
We used the VLBA X-band receiver centered at 8.67~GHz for comparability with the VLA observations presented in \citet[][]{2020ApJ...888...36R}. Our main conclusions are as follows:

\begin{enumerate}
\item We confirmed the presence of compact sources consistent with accretion onto MBHs in 4 out of the 13 VLA sources. However, these 4 sources are the furthest from their associated galaxy photo-centers (between 2\farcs5 and 5\arcsec), a result that is significant at the $p=0.0014$ level, and with distances consistent with expectations for background AGNs. 

\item The non-detection of VLBA counterparts for 9 out of the 13 VLA sources indicates that a significant fraction of the VLA flux density is extended beyond the largest angular scale detectable with the VLBA observations (${\sim}30$ mas). While this phenomenon is commonplace in AGNs, where the emission on arcsecond scales may be dominated by outflows, winds, or lobes, it is also consistent with expectations for star formation-related processes, and indeed the VLBA observations set an upper limit on the brightness temperatures that is consistent with this scenario.

\item However, by extending the SFR analysis performed in \citet[][]{2020ApJ...888...36R}, we found that $\sim5$ out of the 9 VLBA non-detections have VLA sources that are likely too luminous for their host galaxies' SFRs, favoring either the wandering accreting MBH scenario or background AGN. Without spectroscopic confirmation, however, we cannot rule out that there are remaining background AGNs in these objects.

\item Two of the VLBA non-detections with radio luminosities consistent with expectations from star formation were previously confirmed to be AGNs using optical spectroscopy and X-ray data, so their radio emission may indeed be AGN-related. One is a mid-IR AGN, which may lead to an over-estimate of the host galaxy's SFR, increasing this likelihood. These two AGNs are notably very close to, or consistent with, their host-galaxy photo-centers, making their confirmation as AGNs less surprising.

\end{enumerate}

Thus far, the nature of six of the 13 radio-selected AGN candidates in dwarf galaxies presented by \citet{2020ApJ...888...36R} have been determined. Four sources are likely background AGNs and two objects are confirmed to be active MBHs near the centers of their dwarf host galaxies based on optical spectroscopy. The origin of the remaining seven VLA sources from \citet{2020ApJ...888...36R}, with radio-optical position offsets between ${\sim}0\arcsec$ and 2\farcs5, have yet to be determined, although $\sim5$ of them appear to be too luminous to be star formation-related processes such as \ion{H}{2} regions or SN/SNe. Follow-up multi-wavelength observations, such as with the Hubble Space Telescope, are being obtained to better understand the nature of these sources.

\begin{acknowledgements}
This research made use of Astropy,\footnote{\url{http://www.astropy.org}} a community-developed core Python package for Astronomy \citep{2013A&A...558A..33A, 2018AJ....156..123A}. The National Radio Astronomy Observatory is a facility of the National Science Foundation operated under cooperative agreement by Associated Universities, Inc. The authors acknowledge use of the Very Long Baseline Array under the U.S.\ Naval Observatory’s time allocation. 
\end{acknowledgements}

\facilities{GALEX, WISE, VLA, VLBA}

\software{\textsc{aips} \citep{1996ASPC..101...37V}, Astropy \citep{2013A&A...558A..33A, 2018AJ....156..123A}}

\bibliography{vlbadata}
\bibliographystyle{aasjournal}

\end{document}